\documentclass[aps,pre,superscriptaddress,twocolumn,amsmath,amssymb,showpacs]{revtex4}
\usepackage{graphicx}% Include figure files
\usepackage{dcolumn}% Align table columns on decimal point
\usepackage{bm}% bold math
\usepackage{color}

\newcommand{\BE}{\begin{equation}}
\newcommand{\EE}{\end{equation}}
\newcommand{\BA}{\begin{eqnarray}}
\newcommand{\EA}{\end{eqnarray}}
\newcommand{\bx}{{\bf x}}
\newcommand{\by}{{\bf y}}
\newcommand{\bk}{{\bf k}}
\newcommand{\hG}{{\hat G}}
\begin{document}

%%%%%%%%%%%%%%%%%%%%%%%%%%%%%%%%%%%%%%%%

\title{Clustering determines the survivor for competing Brownian and L\'evy walkers}

%%%%%%%%%%%%%%%%%%%%%%%%%%%%%%%%%%%%%%%%

\author{Els Heinsalu}
\affiliation{Niels Bohr International Academy, Niels Bohr Institute, Blegdamsvej 17,
DK-2100 Copenhagen, Denmark}
\affiliation{National Institute of Chemical Physics and Biophysics, R\"avala 10, 15042
Tallinn, Estonia}

\author{Emilio Hern\'andez-Garcia}
\affiliation{IFISC, Instituto de F\'isica Interdisciplinar y Sistemas Complejos (CSIC-UIB),
E-07122 Palma de Mallorca, Spain}

\author{Crist\'obal L\'opez}
\affiliation{IFISC, Instituto de F\'isica Interdisciplinar y Sistemas Complejos (CSIC-UIB),
E-07122 Palma de Mallorca, Spain}

\date{\today}

%%%%%%%%%%%%%%%%%%%%%%%%%%%%%%%%%%%%%%%%

\begin{abstract}
The competition between two ecologically similar species that use the same resources and differ from each other only in the type of spatial motion they undergo is studied. 
The latter is assumed to be described either by Brownian motion or L\'evy flights. 
Competition is taken into account by assuming that individuals reproduce in a density-dependent fashion. 
It is observed that no influence of the type of motion occurs when the two species are in a well-mixed unstructured state.
However, as soon as the species develop spatial clustering, the one forming more concentrated clusters gets a competitive advantage and eliminates the other. 
Similar competitive advantage would occur between walkers of the same type but with different diffusivities if this leads also to different clustering.
Coexistence of both species is also possible under certain conditions.
\end{abstract}

%%%%%%%%%%%%%%%%%%%%%%%%%%%%%%%%%%%%%%%%

%\keywords{...}

\maketitle

The basic ecological factors determining the quantity and distribution of organisms are the reproduction and death processes, which are influenced by the competition for resources, and the dispersal of individuals \cite{Okubo2001,Murray2002}. 
In statistical physics such systems can be addressed using interacting particle models. 
On the basis of the organisms dispersal, interacting Brownian and L\'evy bug models have been proposed \cite{Hernandez2004,Heinsalu2010}. 
In these models the competition is taken into account assuming that demographic processes depend on population density. 
For appropriate parameters, a salient property of these models is the formation of a spatially periodic clustering of individuals.

Most studies addressing the role of dispersal in population dynamics have focussed on the efficiency of foraging or avoiding predation. 
Examples are searching strategies \cite{Benichou2011,James2011} that have revealed the advantage of L\'evy motion with respect to the Brownian one under certain conditions. 
Other studies have addressed collective motion \cite{Vicsek2012,Romanczuk2012}, patchy characteristics of organisms distributions \cite{Zhang1990,Shnerb2000,Young2001,Martin2003,Shnerb2004,Clerc2005,Cecconi2007,reichenbach2008,Houchmandzadeh2009,Neufeld2010}, or the role of demographic fluctuations \cite{Ramos2008,Brigatti,Butler2009,Bonachela2012,Olla2012,Rogers2012a,Rogers2012b}.
An open question is whether the type of motion can enhance the survival probability of competing species.

To understand the factors leading to the extinction, survival, or coexistence of competing species, is a main aim in population ecology. 
It has been shown that the formation of patches is one of the key promoters for species diversity \cite{Reichenbach2007,reichenbach2008,Hassel1994}. 
Cluster and patch formation, with its influence on competition processes, is affected by the dispersal of individuals \cite{Heinsalu2010,Heinsalu2012,Hernandez2004,Reichenbach2007,reichenbach2008,Kerr2002,Harrison2001,Hanski1999}.

In this paper we address the interplay between dispersal and interactions based on competition. 
We consider a system in which initially half of the organisms are characterized by Brownian motion whereas the other half by L\'evy flights, being otherwise identical. 
For example, one can think of the foraging behavior of two types of microorganisms, competing for the same resource and whose spatial motion is consistent either with Brownian or L\'evy random walks \cite{Houchmandzadeh2008,Matthaus2009,Matthaus2011}. 
In particular, the motion of {\it Escherichia coli} is believed to correspond to Brownian diffusion, however, experiments have indicated that some subpopulations perform L\'evy walks \cite{Matthaus2009,Matthaus2011}.
The objective of the present work is to determine which of the two species survives, and if coexistence is possible. 
Our main result is that survival is mediated by the clustering, so that forming stronger clusters provides better chances for survival.
Species coexistence is also observed under certain conditions.

Although we concentrate on the comparison between species undergoing Brownian and L\'{e}vy motion, similar conclusions can be drawn when comparing two species whose dispersion is described by the same type of motion with different diffusion coefficients \cite{Mitchell1991,Butenko2012}.

{\it Model.} --- We consider a system consisting initially of $N_0 = 1000$ organisms, modeled as point-like particles (bugs/walkers). 
Half of them are Brownian random walkers characterized by a diffusion coefficient, $\kappa$, and the other half L\'evy random walkers characterized by a generalized diffusion coefficient, $\kappa_\mu$. 
Brownian walkers perform Gaussianly distributed jumps so that the variance of the displacement of each individual grows proportionally to $\kappa t$. 
L\'{e}vy organisms perform jumps of length $l$ sampled from a L\'evy-type probability density that for large $l$ behaves as $\varphi_\mu (l) \approx|l|^{-\mu - 1}$, with $\mu \in (0, 2)$ being the anomalous exponent; the smaller the value of $\mu$ the more anomalous the random walk.
The variance of the displacement is divergent, but one can identify from moments of sufficiently low order a growing displacement that scales as $x\sim (\kappa_\mu t)^{1/\mu}$.

Besides performing the two-dimensional continuous time random walk, the individuals die following a Poisson process of constant rate $r_{d0}$ and reproduce at rate
\begin{equation}
r^i_b = \mathrm{max} \left( 0, r_{b0} - \alpha N_R^i \right) \, , \label{r-birth}
\end{equation}
i.e., the reproduction probability of an individual $i$ depends on the number of its neighbors, $N_R^i$, that are at a distance smaller than $R$ ($R \ll L$). 
It is assumed that $\alpha > 0$, i.e., the organisms interact in a competitive way. 
Newborns are placed at the same position as the parent, leading to reproductive correlations, and use the same type of motion.

The system is simulated through the Gillespie algorithm as described in Ref.~\cite{Heinsalu2012}. 
Throughout the paper we assume that $r_{d0} = 0.1$, $r_{b0} = 1$, $\alpha = 0.02$, $R = 0.1$, and  $\mu = 1$. 
The only parameters that we are going to vary are $\kappa$ and $\kappa_\mu$.

\begin{figure}[t] \centering
\includegraphics[width=8.5cm]{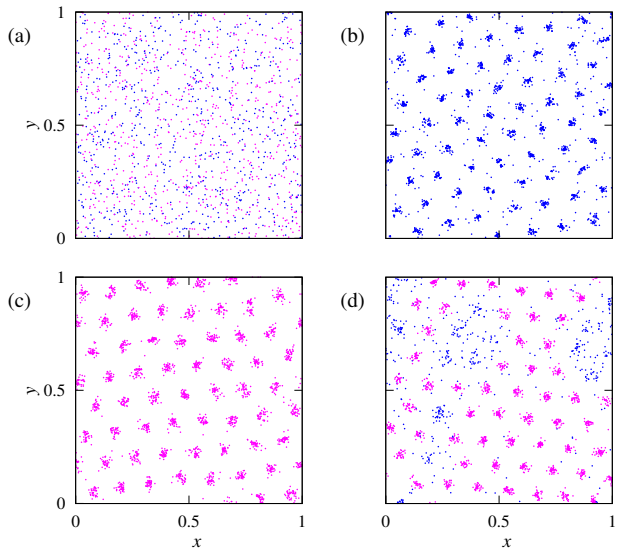}
\caption{
(Color online) Spatial configurations of L\'{e}vy (blue) and Brownian (magenta) organisms at long times:
(a) coexistence without clustering at time $t = 1500$, $\kappa = 1$ and $\kappa_\mu = 1$ (cf. Fig.~\ref{Fig-NvsT}-a for the population sizes);
(b) the L\'{e}vy bugs with $\kappa_\mu = 4 \times 10^{-4}$ have won the Brownian ones with $\kappa = 10^{-5}$;
(c) the Brownian bugs with $\kappa = 8 \times 10^{-6}$ have won the L\'{e}vy ones with $\kappa_\mu = 4 \times 10^{-2}$;
(d) coexistence with clustering at time $t = 173000$, $\kappa = 6 \times 10^{-6}$ and $\kappa_\mu = 4 \times 10^{-3}$ (cf. Fig.~\ref{Fig-NvsT}-b for the population sizes). 
}
\label{Fig-4configs}
\end{figure}
\begin{figure}[t]
\centering
\includegraphics[width=7cm]{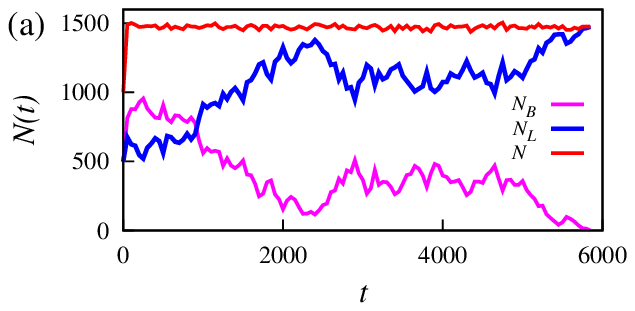}\\
\includegraphics[width=7cm]{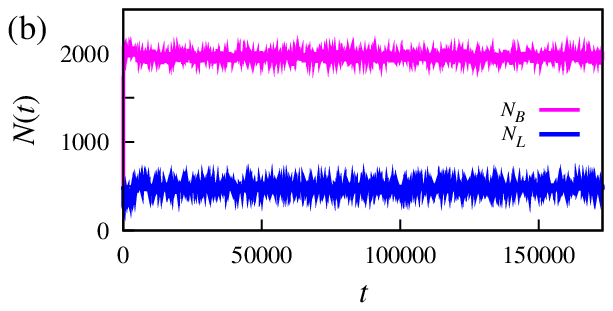}
\caption{
(Color online) Time evolution of the population sizes of the L\'{e}vy and Brownian walkers, $N_B$ and $N_L$, in the two possible cases of coexistence:
(a) large diffusion coefficients when no clustering occurs, $\kappa_{\mu}=1$ and $\kappa=1$ (the same system as in Fig.~\ref{Fig-4configs}-a);
(b) small diffusion coefficients leading to the clustering, $\kappa_\mu = 4 \times 10^{-3}$ and $\kappa = 6 \times 10^{-6}$ (the same system as in Fig.~\ref{Fig-4configs}-d).
}
\label{Fig-NvsT}
\end{figure}

{\it Results.} --- For large values of $\kappa$ and $\kappa_\mu$ the walkers appear to be distributed in an unstructured way, see Fig.~\ref{Fig-4configs}-a. 
Local fluctuations occur around an homogeneous mean, but there is no stable pattern forming. 
The population sizes of Brownian and L\'{e}vy walkers, $N_B$ and $N_L$, widely fluctuate in antiphase, but the total number of individuals, $N = N_B + N_L$, remains quasi-constant, Fig.~\ref{Fig-NvsT}-a. 
The ensemble average reveals the coexistence of the two species. 
In fact, when both species are highly diffusive, the two types of walkers become well mixed, so that there is no difference in the neighborhood seen by the individuals of the different species. 
From the point of view of the interactions the two species become equivalent, and neutral fluctuations are expected from the randomness of the reproduction-death process. 
However, in single realizations the large fluctuations bring one of the species into extinction at long times (see Fig.~\ref{Fig-NvsT}-a) with no possible recovery as individuals can only arise from ancestors of the same type.

Decreasing $\kappa$ or $\kappa_\mu$ the situation changes qualitatively: the corresponding walkers begin to cluster in groups that form a quasi-hexagonal pattern. 
The clusters are different for the two species \cite{Heinsalu2010,Heinsalu2012}, i.e., there is a clear segregation and the competition interaction is felt differently by the two types of organisms, leading eventually to the extinction of one of the species (see Figs.~\ref{Fig-4configs}-b and -c).

For a restricted range of parameters, however, we observe coexistence of L\'{e}vy and Brownian bugs also in the case of low diffusion coefficients. 
In Fig.~\ref{Fig-4configs}-d most of the clusters consist of Brownian walkers, but some clusters in the pattern are replaced by the L\'{e}vy ones.
The population sizes of Brownian and L\'evy walkers, $N_B$ and $N_L$, reach a stationary value rather fast and fluctuate only slightly around it, see Fig.~\ref{Fig-NvsT}-b. 
The average population sizes are constant over a long time, indicating the coexistence of the two species. 
In this case coexistence and segregation happen simultaneously, differently from the mixed up situation for large $\kappa$ and $\kappa_\mu$. 
The transition from homogeneous distribution to the clustered state is similar to the instabilities in the case of single species systems \cite{Hernandez2004,Heinsalu2010,Heinsalu2012}.

\begin{figure}[t]
\centering
\includegraphics[width=7.5cm]{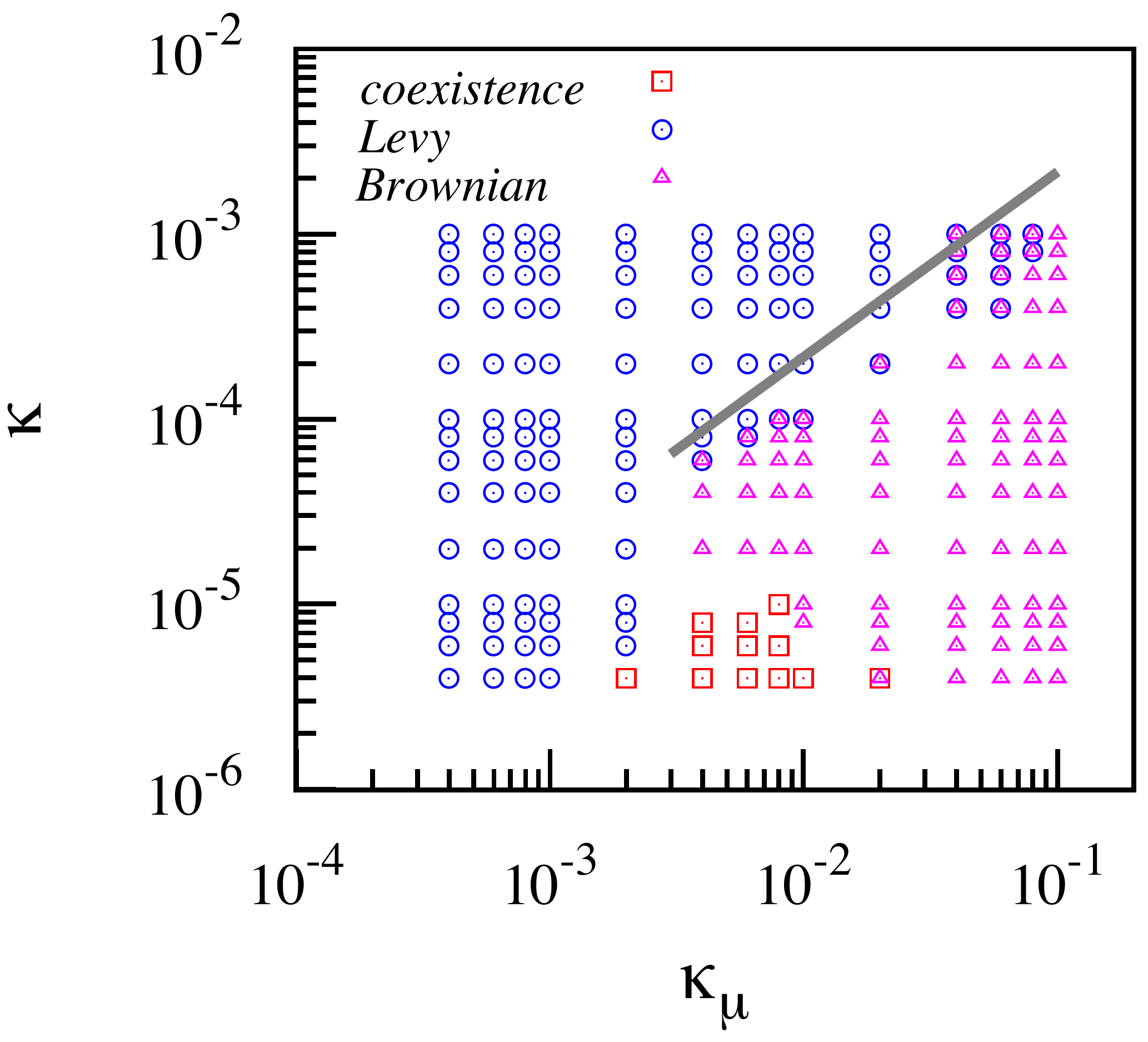}
\caption{
(Color online) Competition between Brownian and L\'evy walkers.
Depending on the values of $\kappa$ and $\kappa_\mu$ either Brownian or L\'evy walkers win, or coexistence occurs.
Each point reflects the outcome of 25 realizations. 
The solid line presents the separation line $\kappa = 0.0217 \kappa_\mu$ provided by the mean-field description.
}
\label{Fig-Phase}
\end{figure}

An overview of the outcome of the competition between Brownian and L\'evy walkers depending on the values of $\kappa$ and $\kappa_\mu$ ($\mu = 1$) is given in Fig.~\ref{Fig-Phase}. 
In the chosen range of $\kappa$, for a fixed value of $\kappa_\mu$, three situations can occur: 1) at small values of $\kappa_\mu$ L\'evy walkers win; 2) at large $\kappa_\mu$ Brownian walkers win; 3) at intermediate values of $\kappa_\mu$, depending on the value of $\kappa$, L\'evy or Brownian walkers win, or coexistence occurs.
In the transition from one regime to the other different runs can lead to different results. 
For other values of  $\mu$ the pictures are similar, with the difference that decreasing $\mu$ transitions are shifted to higher and increasing $\mu$ to smaller values of $\kappa_\mu$. 
Varying other parameters influences the results similarly as discussed in Refs.~\cite{Hernandez2004,Heinsalu2012}.

{\it Mean-field description.} --- In order to gain some understanding of the outcome of the competition process, let us analyze a mean-field description of the system.
Denoting the local densities of Brownian and L\'evy walkers by $\rho_B(\bx,t)$ and by $\rho_L(\bx,t)$, standard arguments, in which statistical fluctuations are neglected, lead to the following dynamics \cite{Hernandez2004,Lopez2004,HernandezPA2005,Heinsalu2010,Heinsalu2012}:
\BE
\begin{aligned}
\frac{\mathrm{\partial} \rho_B(\bx,t)}{\mathrm{\partial} t} &=& M(\bx,t) \rho_B(\bx,t)  + \kappa \nabla^2 \rho_B(\bx,t) \, ,\\
\frac{\mathrm{\partial} \rho_L(\bx,t)}{\mathrm{\partial} t} &=& M(\bx,t) \rho_L(\bx,t)  + \kappa_\mu \nabla^\mu \rho_L(\bx,t) \, .
\label{2eqs}
\end{aligned}
\EE
Here $M(\bx,t) \equiv \beta - G_\bx* \left( \rho_B+\rho_L\right)$, with the net linear growth rate $\beta = r_{b0} - r_{d0}$ . 
$\nabla^\mu$ stands for the fractional derivative of order $\mu$ associated to the L\'{e}vy process \cite{Metzler2000,Klages2008}. 
The symbol $G_\bx*$ denotes the convolution product with a kernel $G(\bx)$, i.e., $G_\bx* f \equiv \int \mathrm{d} \by \, G(\bx-\by) f(\by)$, where the integration is over all system domain. 
Interactions enter the dynamics via Eq.~(\ref{r-birth}) so that  $G(\bx) =\alpha$ if $|\bx|<R$, and $G(\bx) =0$ elsewhere. 
The Fourier transform of $G(\bx)$ is $\hG_\bk = \int \mathrm{d} \bx \, e^{i \bk \cdot \bx} G(\bx)$, and $\hG_0 \equiv \hG_{\bk = {\bf 0}} = \int \mathrm{d} \bx \, G(\bx)$, so that for our two-dimensional case these functions become: $\hG_\bk = 2 \alpha \pi R^2 J_1 (kR) / (kR)$, with $J_1$ being the first-order Bessel function, $\hG_0 = \alpha \pi R^2$, and $k = |\bk|$.
Equations~(\ref{2eqs}) neglect the max condition in Eq.~(\ref{r-birth}), which is not very relevant for the present parameter values (see Ref.~\cite{Hernandez2004}).

We first look for the spatially homogeneous solutions of Eqs.~(\ref{2eqs}). 
In this case the spatial derivatives vanish and there is no difference between the dynamics of the two species. 
There exists a family of steady homogeneous solutions satisfying the condition $\rho_B + \rho_L = \beta / \hat G_0$. 
Thus, we can describe the members of such a family in terms of a parameter $a \in [-\beta / (2 \hG_0),\beta / (2 \hG_0)]$:
\BE
\rho_B^0 = \frac{\beta}{2 \hG_0} + a \, , ~~~ \rho_L^0 = \frac{\beta}{2 \hG_0} - a \, .
\label{homogeneous}
\EE
The upper boundary of this family ($a =\beta / (2 \hG_0)$) corresponds to the pure Brownian, whereas the lower boundary ($a = - \beta / (2 \hG_0)$) to the pure L\'{e}vy population. 
Intermediate values of $a$ parameterize different degrees of homogeneous coexistence.

To demonstrate that this homogeneous family is stable for sufficiently high values of $\kappa$ and $\kappa_\mu$, we perturb it with harmonic functions and look at the growth rates of such perturbations: $\rho_B(\bx,t) = \rho_B^0 + \delta_B \, e^{\lambda t} e^{i\bk\cdot\bx}$ and $\rho_L(\bx,t) = \rho_L^0 + \delta_L \, e^{\lambda t} e^{i\bk\cdot\bx}$ . Linearizing respect to the small perturbations $\delta_B$ and $\delta_L$, one gets a linear system for which the solvability conditions give a quadratic equation for $\lambda$, with two solutions, $\lambda_\pm$, for each value of $k$ (and fixed model parameters). 
For sufficiently large diffusion coefficients the values of $\lambda_+$ and $\lambda_-$ are negative (except for the zero mode $\lambda_+ (k = 0) = 0$), meaning that any perturbation applied decays (except the neutral ones associated to the zero mode, which transforms one of the homogeneous solutions into another one), and thus any of the homogeneous solutions is stable. 
No persistent pattern appears in the system for large values of $\kappa$ and $\kappa_\mu$.
Notice that for the parameter values used in Figs.~\ref{Fig-4configs}-a and \ref{Fig-NvsT}-a, we have that $\beta = 0.9$ and $\rho_B^0 + \rho_L^0 = \beta / \hG_0 \approx 1433$, in good agreement with the total population size $N$ in the numerical simulation. 
At each instant the system is in one of the homogeneous states described by Eqs.~(\ref{homogeneous}), but with continuous fluctuations in the direction of the neutral mode (equivalent to fluctuations in $a$), transforming one of the homogeneous states into another, due to the random birth-death process.

Decreasing $\kappa$ or $\kappa_\mu$, the growth rate $\lambda_+$ becomes positive at a finite value of $k$. 
A pattern forming instability occurs leading to periodic modulations of the densities with a characteristic periodicity given by $2 \pi / k$, similarly to the cases of a single species \cite{Hernandez2004,Heinsalu2010,Heinsalu2012}.
The instability occurs when
\BE
\kappa k^2 \kappa_\mu k^\mu + \frac{\beta \hG_\bk}{2 \hG_0}(\kappa_\mu k^\mu + \kappa k^2) + a \hG_\bk (\kappa_\mu k^\mu - \kappa k^2) < 0 \ ,
\label{InstabilityCondition}
\EE
which happens first for values of $\bk$ leading to negative values of $\hG_\bk$ and for $a (\kappa_\mu k^\mu - \kappa k^2)
> 0$. 
Due to the linear dependence in $a$, the earliest instability appears for the values of $a$ at the extremes of its definition range, i.e., for $a = - \beta / (2 \hat G_0)$ if $\kappa k^2 >\kappa_\mu k^\mu$ and for $a = \beta / (2 \hat G_0)$ if $\kappa k^2 < \kappa_\mu k^\mu$. 
The unstable mode associated to these instabilities involves only the L\'{e}vy or the Brownian population, respectively, so that the pattern that will grow from the unstable state will contain only that species. 
Once clusters appear in some part, they will dominate the whole system. 
The value of $k$ in the above expressions is the one at which the condition (\ref{InstabilityCondition}) is first achieved, i.e., $k_B \approx 4.77901 / R$ for the Brownian homogeneous background \cite{Hernandez2004}, and $k_L\approx 4.94708 / R$ for the L\'{e}vy homogeneous background ($\mu = 1$) \cite{Heinsalu2010}. 
The associated periodicity, $2 \pi / k_B$ or $2 \pi / k_L$, is between $R$ and $2R$ and the separation line $\kappa k_B^2 = \kappa_\mu k_L^\mu$ in $\kappa$-$\kappa_\mu$ space between the two winning states is given by $\kappa = 0.0217 \kappa_\mu$.
As can be seen from Fig.~\ref{Fig-Phase}, at intermediate values of $\kappa_\mu$ the separation line found from the mean-field description follows rather well the trend of the numerically determined transition.

Thus, the picture emerging from the mean-field description is the following: at large values of $\kappa$ and $\kappa_\mu$ the two types of organisms are essentially the same and coexistence occurs (until a neutral fluctuation eliminates irreversibly one of the species). 
When decreasing $\kappa$ and/or $\kappa_\mu$, mixing becomes less good and different regions of the system may be occupied by different proportions of bug densities that satisfy the condition $\rho_B + \rho_L = \beta / \hat G_0$. 
By further decreasing the diffusion coefficients (or increasing $\beta$ or $R$), some of these regions will encounter an instability. 
The winning competitor is the one for which the diffusive decay rate of the periodic structure, $\kappa k_B^2$ and $\kappa_\mu k_L^\mu$ for the Brownian and L\'{e}vy bugs respectively, is smaller. 
Note that these quantities can also be interpreted as the density fluxes going out of the clusters. 
The type of walker with the highest flux out of the clusters is the one that loses the competition, being the winner the one that stays well concentrated into strong clusters. 
In fact, we have checked that there is a good correspondence between the type of walker that produces the narrowest and strongest clusters in single-species simulations and the winner of the competition when both types of bugs are allowed to interact at the same parameter values.

Using the idea that competition success is attained by the motion leading to the strongest clustering, one would predict that for the values $\kappa_\mu = 4 \times 10^{-3}$ and $\kappa = 4 \times 10^{-6}$ the Brownian walkers win. 
However, as indicated by Fig.~\ref{Fig-Phase}, this is not the case. 
Instead, coexistence of the two species occurs, i.e., during the accessible simulation time we do not observe the extinction of neither species. 
The arguments presented above do not capture the coexistence of the species in the case of clustering. 
There are many ingredients that have been neglected in the mean-field description, in particular, fluctuations.
Another drawback is that the linear analysis of the mean-field theory predicts that the pattern wavenumber is independent of diffusion coefficient, whereas it was noted in numerical simulations of the single species system that there is some dependence on it \cite{Heinsalu2012}. 
However, the conditions for the coexistence are not difficult to understand. 
What happens is that the Brownian walkers form very strong clusters that the L\'evy walkers are not able to invade despite they are able to wander around them. 
On the other hand, due to the extremely low diffusion and the high death rate in the inter-clusters space, the Brownian walkers are not capable to colonize the territories that have been occupied by the L\'evy walkers during the initial cluster formation due to random fluctuations. 
As a result, the situation depicted in Figs.~\ref{Fig-4configs}-d and \ref{Fig-NvsT}-b is observed.

{\it Resume.} --- In this paper we addressed the impact of dispersal on the competition of organisms that are identical in all other aspects. 
It was observed that no competitive advantage occurs when the organisms are in a well-mixed unstructured state. 
But as soon as clustering occurs in at least one of the species, a competitive advantage is manifested in favor of the species with the stronger clustering. 
The mechanism behind this is that respect to the individuals of the species with a larger flux out of the clusters, the individuals of the species forming stronger clusters experience less the high competition occurring in between the clusters and have therefore a higher probability for reproduction, leading to a higher probability of surviving.
Our results agree with the observations made in earlier works that the dispersal has a role in species competition \cite{Reichenbach2007,reichenbach2008,Kerr2002,Harrison2001,Hanski1999}, but through the simple model and the assumption that the species are identical in all the rest we show that dispersal and the associated cluster formation can be the key feature that determines the outcome of the competition.

{Acknowledgements.} --- This work has been supported by the targeted financing project SF0690030s09, Estonian Science Foundation grant no.~9462 (EH), and by Spanish MINECO and FEDER through project FISICOS (FIS2007-60327), ESCOLA (CTM2012-39025-C02-01) and INTENSE@COSYP (FIS2012-30634).

\bibliography{Refs-Heinsalu}

\begin{thebibliography}{40}
\expandafter\ifx\csname natexlab\endcsname\relax\def\natexlab#1{#1}\fi
\expandafter\ifx\csname bibnamefont\endcsname\relax
  \def\bibnamefont#1{#1}\fi
\expandafter\ifx\csname bibfnamefont\endcsname\relax
  \def\bibfnamefont#1{#1}\fi
\expandafter\ifx\csname citenamefont\endcsname\relax
  \def\citenamefont#1{#1}\fi
\expandafter\ifx\csname url\endcsname\relax
  \def\url#1{\texttt{#1}}\fi
\expandafter\ifx\csname urlprefix\endcsname\relax\def\urlprefix{URL }\fi
\providecommand{\bibinfo}[2]{#2}
\providecommand{\eprint}[2][]{\url{#2}}

\bibitem[{\citenamefont{Okubo and Levin}(2001)}]{Okubo2001}
\bibinfo{author}{\bibfnamefont{A.}~\bibnamefont{Okubo}} \bibnamefont{and}
  \bibinfo{author}{\bibfnamefont{S.}~\bibnamefont{Levin}},
  \emph{\bibinfo{title}{Diffusion and Ecological Problems}}
  (\bibinfo{publisher}{Springer}, \bibinfo{address}{New York},
  \bibinfo{year}{2001}), \bibinfo{edition}{2nd} ed.

\bibitem[{\citenamefont{Murray}(2002)}]{Murray2002}
\bibinfo{author}{\bibfnamefont{J.~D.} \bibnamefont{Murray}},
  \emph{\bibinfo{title}{Mathematical Biology I. An Introduction}},
  Interdisciplinary Applied Mathematics, vol 17 (\bibinfo{publisher}{Springer},
  \bibinfo{address}{New York}, \bibinfo{year}{2002}), \bibinfo{edition}{3rd}
  ed.

\bibitem[{\citenamefont{Hern\'andez-Garc\'{\i}a and
  L\'opez}(2004)}]{Hernandez2004}
\bibinfo{author}{\bibfnamefont{E.}~\bibnamefont{Hern\'andez-Garc\'{\i}a}}
  \bibnamefont{and} \bibinfo{author}{\bibfnamefont{C.}~\bibnamefont{L\'opez}},
  \bibinfo{journal}{Phys. Rev.~E} \textbf{\bibinfo{volume}{70}},
  \bibinfo{pages}{016216} (\bibinfo{year}{2004}).

\bibitem[{\citenamefont{Heinsalu et~al.}(2010)\citenamefont{Heinsalu,
  Hern\'andez-Garc\'{\i}a, and L\'opez}}]{Heinsalu2010}
\bibinfo{author}{\bibfnamefont{E.}~\bibnamefont{Heinsalu}},
  \bibinfo{author}{\bibfnamefont{E.}~\bibnamefont{Hern\'andez-Garc\'{\i}a}},
  \bibnamefont{and} \bibinfo{author}{\bibfnamefont{C.}~\bibnamefont{L\'opez}},
  \bibinfo{journal}{Europhys. Lett.} \textbf{\bibinfo{volume}{92}},
  \bibinfo{pages}{40011} (\bibinfo{year}{2010}), \bibinfo{note}{erratum:
  Europhys. Lett. 95, 69902 (2011)}.

\bibitem[{\citenamefont{Benichou et~al.}(2011)\citenamefont{Benichou, Loverdo,
  Moreau, and Voituriez}}]{Benichou2011}
\bibinfo{author}{\bibfnamefont{O.}~\bibnamefont{Benichou}},
  \bibinfo{author}{\bibfnamefont{C.}~\bibnamefont{Loverdo}},
  \bibinfo{author}{\bibfnamefont{M.}~\bibnamefont{Moreau}}, \bibnamefont{and}
  \bibinfo{author}{\bibfnamefont{R.}~\bibnamefont{Voituriez}},
  \bibinfo{journal}{Reviews of Modern Physics} \textbf{\bibinfo{volume}{83}},
  \bibinfo{pages}{81} (\bibinfo{year}{2011}).

\bibitem[{\citenamefont{James et~al.}(2011)\citenamefont{James, Plank, and
  Edwards}}]{James2011}
\bibinfo{author}{\bibfnamefont{A.}~\bibnamefont{James}},
  \bibinfo{author}{\bibfnamefont{M.~J.} \bibnamefont{Plank}}, \bibnamefont{and}
  \bibinfo{author}{\bibfnamefont{A.~M.} \bibnamefont{Edwards}},
  \bibinfo{journal}{J. Royal Soc. Interface} \textbf{\bibinfo{volume}{8}},
  \bibinfo{pages}{1233} (\bibinfo{year}{2011}).

\bibitem[{\citenamefont{Vicsek and Zafiris}(2012)}]{Vicsek2012}
\bibinfo{author}{\bibfnamefont{T.}~\bibnamefont{Vicsek}} \bibnamefont{and}
  \bibinfo{author}{\bibfnamefont{A.}~\bibnamefont{Zafiris}},
  \bibinfo{journal}{Physics Reports} \textbf{\bibinfo{volume}{517}},
  \bibinfo{pages}{71} (\bibinfo{year}{2012}).

\bibitem[{\citenamefont{Romanczuk et~al.}(2012)\citenamefont{Romanczuk, B\"ar,
  Ebeling, Lindner, and Schimansky-Geier}}]{Romanczuk2012}
\bibinfo{author}{\bibfnamefont{P.}~\bibnamefont{Romanczuk}},
  \bibinfo{author}{\bibfnamefont{M.}~\bibnamefont{B\"ar}},
  \bibinfo{author}{\bibfnamefont{W.}~\bibnamefont{Ebeling}},
  \bibinfo{author}{\bibfnamefont{B.}~\bibnamefont{Lindner}}, \bibnamefont{and}
  \bibinfo{author}{\bibfnamefont{L.}~\bibnamefont{Schimansky-Geier}},
  \bibinfo{journal}{Eur. Phys. J. Special Topics}
  \textbf{\bibinfo{volume}{202}}, \bibinfo{pages}{1} (\bibinfo{year}{2012}).

\bibitem[{\citenamefont{Zhang et~al.}(1990)\citenamefont{Zhang, Serva, and
  Polikarpov}}]{Zhang1990}
\bibinfo{author}{\bibfnamefont{Y.-C.} \bibnamefont{Zhang}},
  \bibinfo{author}{\bibfnamefont{M.}~\bibnamefont{Serva}}, \bibnamefont{and}
  \bibinfo{author}{\bibfnamefont{M.}~\bibnamefont{Polikarpov}},
  \bibinfo{journal}{J.~Stat. Phys.} \textbf{\bibinfo{volume}{58}},
  \bibinfo{pages}{849} (\bibinfo{year}{1990}).

\bibitem[{\citenamefont{Shnerb et~al.}(2000)\citenamefont{Shnerb, Louzoun,
  Bettelheim, and Solomon}}]{Shnerb2000}
\bibinfo{author}{\bibfnamefont{N.}~\bibnamefont{Shnerb}},
  \bibinfo{author}{\bibfnamefont{Y.}~\bibnamefont{Louzoun}},
  \bibinfo{author}{\bibfnamefont{E.}~\bibnamefont{Bettelheim}},
  \bibnamefont{and} \bibinfo{author}{\bibfnamefont{S.}~\bibnamefont{Solomon}},
  \bibinfo{journal}{Proceedings of the National Academy of Sciences}
  \textbf{\bibinfo{volume}{97}}, \bibinfo{pages}{10322} (\bibinfo{year}{2000}).

\bibitem[{\citenamefont{Young et~al.}(2001)\citenamefont{Young, Roberts, and
  Stuhne}}]{Young2001}
\bibinfo{author}{\bibfnamefont{W.~R.} \bibnamefont{Young}},
  \bibinfo{author}{\bibfnamefont{A.~J.} \bibnamefont{Roberts}},
  \bibnamefont{and} \bibinfo{author}{\bibfnamefont{G.}~\bibnamefont{Stuhne}},
  \bibinfo{journal}{Nature} \textbf{\bibinfo{volume}{412}},
  \bibinfo{pages}{328} (\bibinfo{year}{2001}).

\bibitem[{\citenamefont{Martin}(2003)}]{Martin2003}
\bibinfo{author}{\bibfnamefont{A.}~\bibnamefont{Martin}},
  \bibinfo{journal}{Progress in Oceanography} \textbf{\bibinfo{volume}{57}},
  \bibinfo{pages}{125} (\bibinfo{year}{2003}).

\bibitem[{\citenamefont{Shnerb}(2004)}]{Shnerb2004}
\bibinfo{author}{\bibfnamefont{N.}~\bibnamefont{Shnerb}},
  \bibinfo{journal}{Phys. Rev. E} \textbf{\bibinfo{volume}{69}},
  \bibinfo{pages}{061917} (\bibinfo{year}{2004}).

\bibitem[{\citenamefont{Clerc et~al.}(2005)\citenamefont{Clerc, Escaff, and
  Kenkre}}]{Clerc2005}
\bibinfo{author}{\bibfnamefont{M.~G.} \bibnamefont{Clerc}},
  \bibinfo{author}{\bibfnamefont{D.}~\bibnamefont{Escaff}}, \bibnamefont{and}
  \bibinfo{author}{\bibfnamefont{V.~M.} \bibnamefont{Kenkre}},
  \bibinfo{journal}{Phys. Rev. E} \textbf{\bibinfo{volume}{72}},
  \bibinfo{pages}{056217} (\bibinfo{year}{2005}).

\bibitem[{\citenamefont{Cecconi et~al.}(2007)\citenamefont{Cecconi, Gonella,
  and Saracco}}]{Cecconi2007}
\bibinfo{author}{\bibfnamefont{F.}~\bibnamefont{Cecconi}},
  \bibinfo{author}{\bibfnamefont{G.}~\bibnamefont{Gonella}}, \bibnamefont{and}
  \bibinfo{author}{\bibfnamefont{G.}~\bibnamefont{Saracco}},
  \bibinfo{journal}{Phys. Rev. E} \textbf{\bibinfo{volume}{75}},
  \bibinfo{pages}{031111} (\bibinfo{year}{2007}).

\bibitem[{\citenamefont{Reichenbach and Frey}(2008)}]{reichenbach2008}
\bibinfo{author}{\bibfnamefont{T.}~\bibnamefont{Reichenbach}} \bibnamefont{and}
  \bibinfo{author}{\bibfnamefont{E.}~\bibnamefont{Frey}},
  \bibinfo{journal}{Phys. Rev. Lett.} \textbf{\bibinfo{volume}{101}},
  \bibinfo{pages}{58102} (\bibinfo{year}{2008}).

\bibitem[{\citenamefont{Houchmandzadeh}(2009)}]{Houchmandzadeh2009}
\bibinfo{author}{\bibfnamefont{B.}~\bibnamefont{Houchmandzadeh}},
  \bibinfo{journal}{Phys. Rev. E} \textbf{\bibinfo{volume}{80}},
  \bibinfo{pages}{051920} (\bibinfo{year}{2009}).

\bibitem[{\citenamefont{Neufeld and
  Hern\'andez-Garc\'{\i}a}(2009)}]{Neufeld2010}
\bibinfo{author}{\bibfnamefont{Z.}~\bibnamefont{Neufeld}} \bibnamefont{and}
  \bibinfo{author}{\bibfnamefont{E.}~\bibnamefont{Hern\'andez-Garc\'{\i}a}},
  \emph{\bibinfo{title}{Chemical and Biological Processes in Fluid Flows: A
  Dynamical Systems Approach}} (\bibinfo{publisher}{Imperial College},
  \bibinfo{address}{London}, \bibinfo{year}{2009}).

\bibitem[{\citenamefont{Ramos et~al.}(2008)\citenamefont{Ramos, L\'opez,
  Hern\'andez-Garc\'{\i}a, and Mu{\~n}oz}}]{Ramos2008}
\bibinfo{author}{\bibfnamefont{F.}~\bibnamefont{Ramos}},
  \bibinfo{author}{\bibfnamefont{C.}~\bibnamefont{L\'opez}},
  \bibinfo{author}{\bibfnamefont{E.}~\bibnamefont{Hern\'andez-Garc\'{\i}a}},
  \bibnamefont{and} \bibinfo{author}{\bibfnamefont{M.~A.}
  \bibnamefont{Mu{\~n}oz}}, \bibinfo{journal}{Phys. Rev.~E}
  \textbf{\bibinfo{volume}{77}}, \bibinfo{pages}{021102}
  (\bibinfo{year}{2008}).

\bibitem[{\citenamefont{Brigatti et~al.}(2008)\citenamefont{Brigatti,
  Schwammle, and Neto}}]{Brigatti}
\bibinfo{author}{\bibfnamefont{E.}~\bibnamefont{Brigatti}},
  \bibinfo{author}{\bibfnamefont{V.}~\bibnamefont{Schwammle}},
  \bibnamefont{and} \bibinfo{author}{\bibfnamefont{M.~A.} \bibnamefont{Neto}},
  \bibinfo{journal}{Phys. Rev.~E} \textbf{\bibinfo{volume}{77}},
  \bibinfo{pages}{021914} (\bibinfo{year}{2008}).

\bibitem[{\citenamefont{Butler and Goldenfeld}(2009)}]{Butler2009}
\bibinfo{author}{\bibfnamefont{T.}~\bibnamefont{Butler}} \bibnamefont{and}
  \bibinfo{author}{\bibfnamefont{N.}~\bibnamefont{Goldenfeld}},
  \bibinfo{journal}{Phys. Rev. E} \textbf{\bibinfo{volume}{80}},
  \bibinfo{pages}{030902} (\bibinfo{year}{2009}).

\bibitem[{\citenamefont{Bonachela et~al.}(2012)\citenamefont{Bonachela, Muñoz,
  and Levin}}]{Bonachela2012}
\bibinfo{author}{\bibfnamefont{J.}~\bibnamefont{Bonachela}},
  \bibinfo{author}{\bibfnamefont{M.}~\bibnamefont{Muñoz}}, \bibnamefont{and}
  \bibinfo{author}{\bibfnamefont{S.}~\bibnamefont{Levin}},
  \bibinfo{journal}{Journal of Statistical Physics}
  \textbf{\bibinfo{volume}{148}}, \bibinfo{pages}{723} (\bibinfo{year}{2012}).

\bibitem[{\citenamefont{Olla}(2012)}]{Olla2012}
\bibinfo{author}{\bibfnamefont{P.}~\bibnamefont{Olla}}, \bibinfo{journal}{Phys.
  Rev. E} \textbf{\bibinfo{volume}{85}}, \bibinfo{pages}{021125}
  (\bibinfo{year}{2012}).

\bibitem[{\citenamefont{Rogers et~al.}(2012{\natexlab{a}})\citenamefont{Rogers,
  McKane, and Rossberg}}]{Rogers2012a}
\bibinfo{author}{\bibfnamefont{T.}~\bibnamefont{Rogers}},
  \bibinfo{author}{\bibfnamefont{A.~J.} \bibnamefont{McKane}},
  \bibnamefont{and} \bibinfo{author}{\bibfnamefont{A.~G.}
  \bibnamefont{Rossberg}}, \bibinfo{journal}{EPL (Europhysics Letters)}
  \textbf{\bibinfo{volume}{97}}, \bibinfo{pages}{40008}
  (\bibinfo{year}{2012}{\natexlab{a}}).

\bibitem[{\citenamefont{Rogers et~al.}(2012{\natexlab{b}})\citenamefont{Rogers,
  McKane, and Rossberg}}]{Rogers2012b}
\bibinfo{author}{\bibfnamefont{T.}~\bibnamefont{Rogers}},
  \bibinfo{author}{\bibfnamefont{A.}~\bibnamefont{McKane}}, \bibnamefont{and}
  \bibinfo{author}{\bibfnamefont{A.~G.} \bibnamefont{Rossberg}},
  \bibinfo{journal}{Physical Biology} \textbf{\bibinfo{volume}{9}},
  \bibinfo{pages}{066002} (\bibinfo{year}{2012}{\natexlab{b}}).

\bibitem[{\citenamefont{Reichenbach et~al.}(2007)\citenamefont{Reichenbach,
  Mobilia, and Frey}}]{Reichenbach2007}
\bibinfo{author}{\bibfnamefont{T.}~\bibnamefont{Reichenbach}},
  \bibinfo{author}{\bibfnamefont{M.}~\bibnamefont{Mobilia}}, \bibnamefont{and}
  \bibinfo{author}{\bibfnamefont{E.}~\bibnamefont{Frey}},
  \bibinfo{journal}{Nature} \textbf{\bibinfo{volume}{448}},
  \bibinfo{pages}{1046} (\bibinfo{year}{2007}).

\bibitem[{\citenamefont{Hassell et~al.}(1994)\citenamefont{Hassell, Comins, and
  May}}]{Hassel1994}
\bibinfo{author}{\bibfnamefont{M.~P.} \bibnamefont{Hassell}},
  \bibinfo{author}{\bibfnamefont{H.~N.} \bibnamefont{Comins}},
  \bibnamefont{and} \bibinfo{author}{\bibfnamefont{R.~M.} \bibnamefont{May}},
  \bibinfo{journal}{Nature} \textbf{\bibinfo{volume}{370}},
  \bibinfo{pages}{290} (\bibinfo{year}{1994}).

\bibitem[{\citenamefont{Heinsalu et~al.}(2012)\citenamefont{Heinsalu,
  Hern\'andez-Garc\'{\i}a, and L\'opez}}]{Heinsalu2012}
\bibinfo{author}{\bibfnamefont{E.}~\bibnamefont{Heinsalu}},
  \bibinfo{author}{\bibfnamefont{E.}~\bibnamefont{Hern\'andez-Garc\'{\i}a}},
  \bibnamefont{and} \bibinfo{author}{\bibfnamefont{C.}~\bibnamefont{L\'opez}},
  \bibinfo{journal}{Phys. Rev.~E} \textbf{\bibinfo{volume}{85}},
  \bibinfo{pages}{041105} (\bibinfo{year}{2012}).

\bibitem[{\citenamefont{Kerr et~al.}(2002)\citenamefont{Kerr, Riley, Feldman,
  and Bohannan}}]{Kerr2002}
\bibinfo{author}{\bibfnamefont{B.}~\bibnamefont{Kerr}},
  \bibinfo{author}{\bibfnamefont{M.~A.} \bibnamefont{Riley}},
  \bibinfo{author}{\bibfnamefont{M.~W.} \bibnamefont{Feldman}},
  \bibnamefont{and} \bibinfo{author}{\bibfnamefont{B.~J.~M.}
  \bibnamefont{Bohannan}}, \bibinfo{journal}{Nature}
  \textbf{\bibinfo{volume}{418}}, \bibinfo{pages}{171} (\bibinfo{year}{2002}).

\bibitem[{\citenamefont{Harrison et~al.}(2001)\citenamefont{Harrison, Lai, and
  Holt}}]{Harrison2001}
\bibinfo{author}{\bibfnamefont{M.~A.} \bibnamefont{Harrison}},
  \bibinfo{author}{\bibfnamefont{Y.-C.} \bibnamefont{Lai}}, \bibnamefont{and}
  \bibinfo{author}{\bibfnamefont{R.~D.} \bibnamefont{Holt}},
  \bibinfo{journal}{Phys. Rev.~E} \textbf{\bibinfo{volume}{63}},
  \bibinfo{pages}{051905} (\bibinfo{year}{2001}).

\bibitem[{\citenamefont{Hanski}(1999)}]{Hanski1999}
\bibinfo{author}{\bibfnamefont{I.}~\bibnamefont{Hanski}},
  \emph{\bibinfo{title}{{M}etapopulation {E}cology}}
  (\bibinfo{publisher}{Oxford University Press, Oxford}, \bibinfo{year}{1999}).

\bibitem[{\citenamefont{Houchmandzadeh}(2008)}]{Houchmandzadeh2008}
\bibinfo{author}{\bibfnamefont{B.}~\bibnamefont{Houchmandzadeh}},
  \bibinfo{journal}{Phys. Rev. Lett.} \textbf{\bibinfo{volume}{101}},
  \bibinfo{pages}{078103} (\bibinfo{year}{2008}).

\bibitem[{\citenamefont{Matth{\"a}us et~al.}(2009)\citenamefont{Matth{\"a}us,
  Jagodi\v{c}, and Dobnikar}}]{Matthaus2009}
\bibinfo{author}{\bibfnamefont{F.}~\bibnamefont{Matth{\"a}us}},
  \bibinfo{author}{\bibfnamefont{M.}~\bibnamefont{Jagodi\v{c}}},
  \bibnamefont{and} \bibinfo{author}{\bibfnamefont{J.}~\bibnamefont{Dobnikar}},
  \bibinfo{journal}{Biophysical Journal} \textbf{\bibinfo{volume}{97}},
  \bibinfo{pages}{946 } (\bibinfo{year}{2009}).

\bibitem[{\citenamefont{Matth{\"a}us et~al.}(2011)\citenamefont{Matth{\"a}us,
  Mommer, Curk, and Dobnikar}}]{Matthaus2011}
\bibinfo{author}{\bibfnamefont{F.}~\bibnamefont{Matth{\"a}us}},
  \bibinfo{author}{\bibfnamefont{M.~S.} \bibnamefont{Mommer}},
  \bibinfo{author}{\bibfnamefont{T.}~\bibnamefont{Curk}}, \bibnamefont{and}
  \bibinfo{author}{\bibfnamefont{J.}~\bibnamefont{Dobnikar}},
  \bibinfo{journal}{PLoS ONE} \textbf{\bibinfo{volume}{6}},
  \bibinfo{pages}{e18623} (\bibinfo{year}{2011}).

\bibitem[{\citenamefont{Mitchell}(1991)}]{Mitchell1991}
\bibinfo{author}{\bibfnamefont{J.~G.} \bibnamefont{Mitchell}},
  \bibinfo{journal}{Microb. Ecol.} \textbf{\bibinfo{volume}{21}},
  \bibinfo{pages}{227} (\bibinfo{year}{1991}).

\bibitem[{\citenamefont{Butenko et~al.}(2012)\citenamefont{Butenko, Mogilko,
  Amitai, Pokroy, and Sloutskin}}]{Butenko2012}
\bibinfo{author}{\bibfnamefont{A.~V.} \bibnamefont{Butenko}},
  \bibinfo{author}{\bibfnamefont{E.}~\bibnamefont{Mogilko}},
  \bibinfo{author}{\bibfnamefont{L.}~\bibnamefont{Amitai}},
  \bibinfo{author}{\bibfnamefont{B.}~\bibnamefont{Pokroy}}, \bibnamefont{and}
  \bibinfo{author}{\bibfnamefont{E.}~\bibnamefont{Sloutskin}},
  \bibinfo{journal}{Langmuir} \textbf{\bibinfo{volume}{28}},
  \bibinfo{pages}{12941} (\bibinfo{year}{2012}).

\bibitem[{\citenamefont{L\'opez and Hern\'andez-Garc\'{\i}a}(2004)}]{Lopez2004}
\bibinfo{author}{\bibfnamefont{C.}~\bibnamefont{L\'opez}} \bibnamefont{and}
  \bibinfo{author}{\bibfnamefont{E.}~\bibnamefont{Hern\'andez-Garc\'{\i}a}},
  \bibinfo{journal}{Physica D} \textbf{\bibinfo{volume}{199}},
  \bibinfo{pages}{223} (\bibinfo{year}{2004}).

\bibitem[{\citenamefont{Hern\'andez-Garc\'{\i}a and
  L\'opez}(2005)}]{HernandezPA2005}
\bibinfo{author}{\bibfnamefont{E.}~\bibnamefont{Hern\'andez-Garc\'{\i}a}}
  \bibnamefont{and} \bibinfo{author}{\bibfnamefont{C.}~\bibnamefont{L\'opez}},
  \bibinfo{journal}{Physica A} \textbf{\bibinfo{volume}{356}},
  \bibinfo{pages}{95 } (\bibinfo{year}{2005}).

\bibitem[{\citenamefont{Metzler and Klafter}(2000)}]{Metzler2000}
\bibinfo{author}{\bibfnamefont{R.}~\bibnamefont{Metzler}} \bibnamefont{and}
  \bibinfo{author}{\bibfnamefont{J.}~\bibnamefont{Klafter}},
  \bibinfo{journal}{Phys. Rep.} \textbf{\bibinfo{volume}{339}},
  \bibinfo{pages}{1} (\bibinfo{year}{2000}).

\bibitem[{\citenamefont{Klages et~al.}(2008)\citenamefont{Klages, Radons, and
  Sokolov}}]{Klages2008}
\bibinfo{author}{\bibfnamefont{R.}~\bibnamefont{Klages}},
  \bibinfo{author}{\bibfnamefont{G.}~\bibnamefont{Radons}}, \bibnamefont{and}
  \bibinfo{author}{\bibfnamefont{I.~M.} \bibnamefont{Sokolov}},
  \emph{\bibinfo{title}{{A}nomalous {T}ransport: {F}oundations and
  {A}pplications}} (\bibinfo{publisher}{Wiley-VCH}, \bibinfo{year}{2008}).

\end{thebibliography}

\end{document}